\documentclass[prl, twocolumn, showpacs, preprintnumbers, amsmath, amssymb, superscriptaddress, aps]{revtex4}
\usepackage{graphicx}
\usepackage{dcolumn}
\usepackage{color}
\usepackage{bm}
\def\he4{$^4$He}
\def\Am3{\AA$^{-3}$}
\def\beq{\begin{equation}}
\def\eeq{\end{equation}}
\begin{document}

\author{L. Pollet}
\affiliation{Theoretische Physik, ETH Z\"urich, CH-8093 Z\"urich, Switzerland}

\author{M. Boninsegni}
\affiliation{Department of Physics, University of Alberta,
Edmonton, Alberta T6G 2J1, Canada}

\author{A.B. Kuklov}
\affiliation{Department of Engineering Science and Physics,
CUNY, Staten Island, NY 10314, USA}

\author{N.V. Prokof'ev}
\affiliation{Theoretische Physik, ETH Z\"urich, CH-8093 Z\"urich, Switzerland}
\affiliation{Department of Physics, University of Massachusetts,
Amherst, MA 01003, USA}
\affiliation{Russian Research Center ``Kurchatov Institute'',
123182 Moscow, Russia}

\author{B.V. Svistunov}
\affiliation{Department of Physics, University of Massachusetts,
Amherst, MA 01003, USA}
\affiliation{Russian Research Center ``Kurchatov Institute'',
123182 Moscow, Russia}

\author{M. Troyer}
\affiliation{Theoretische Physik, ETH Z\"urich, CH-8093 Z\"urich, Switzerland}

\title{Local stress and superfluid properties of solid \he4 }

\date{\today}
\begin{abstract}
More than half a century ago Penrose asked~\cite{Penrose51}: are the superfluid
and solid state of matter  mutually exclusive or do there exist ``supersolid" materials where
the atoms form a regular lattice and simultaneously flow without friction? Recent experiments provide evidence that supersolid behavior indeed exists in \he4 -- the most quantum material known in Nature. 
In this paper we show that  large local strain in the vicinity of crystalline defects is the origin of
supersolidity in \he4.
Although ideal crystals of \he4
are not supersolid, the gap for vacancy creation closes when
applying a moderate stress. 
While a homogeneous system simply becomes unstable at this point, the stressed core of crystalline defects (dislocations and grain boundaries) undergoes a radical transformation and can become superfluid.
\end{abstract}
\maketitle

The first microscopic mechanism of supersolidity was proposed by Andreev, Lifshitz \cite{Andreev69} and Chester \cite{Chester70} arguing that a  dilute gas of vacancies
could lower the energy of an ideal helium crystal by quantum mechanically delocalizing vacancies. At low enough temperature the vacancies then undergo Bose-Einstein condensation and give rise to a supersolid -- and indeed vacancies are required for supersolidity \cite{PS05}.
Although the detection of frictionless mass flow was anticipated to be straightforward,
all experimental attempts failed for over thirty years \cite{meisel} until Kim and Chan reported the observation of an unexpected drop of the
torsional oscillator period at low temperatures \cite{KCNature, KCScience}.
Since the period is proportional to the square root of the moment of inertia,
the origin of the period shift was attributed to the supersolid decoupling of the \he4 solid mass from
the rotating walls. Several groups have now confirmed the basic effect
\cite{Rittner06,Aoki,Shirahama,Kubota} but it is still unclear what physics lies
behind these observations, both theoretically and experimentally \cite{Balibar_review}.

First-principle calculations \cite{Boninsegni06a, Clark06}
rule out the possibility of supersolidity in ideal equilibrium crystals
of \he4. On the experimental side, no steady flow of mass through the solid phase was
detected in several setups \cite{Greywall77a,Day}. It is believed that the flow detected
at the melting curve in Ref. \cite{Balibar} was along liquid channels which form on the cell
walls in contact with the grain boundaries, i.e. it was not related to the supersolid phenomenon
\cite{Balibar_review}.

The situation has changed radically with the recent observation of the liquid pressure
equilibration by flow through a solid in the ``UMass sandwich" experiment
reported by Ray and Hallock \cite{Ray} away from the melting curve. The dependence of the flow on the pressure difference was characteristic of critical superflow and inconsistent with the flow of a viscous liquid.
In this work, we explain that the origin of
superfluidity in solid \he4 is large local strain in the vicinity of crystalline defects.
Superfluidity along dislocations or grain boundaries can quantitatively account for the
superflow through \he4 crystals observed by Ray and Hallock \cite{Ray}.

In the past several authors have speculated about the possibility of the stress-induced supersolidity \cite{Schevchenko,Dorsey06,Toner08}, especially under
hydrostatic decompression since quantum effects and vacancy delocalization are expected
to increase at lower densities.  The phenomenology missed the attraction between 
vacancies which destabilizes a dilute homogeneous gas of vacancies. Central questions were left unanswered : what type and strength of stress is needed to close the insulating gap
in \he4 and is this stress realistic?

Our approach to this inherently strongly correlated problem is numerical. Feynman's
path-integral formulation of quantum mechanics allows an exact mapping between the \he4
system and a system of world lines in four-dimensional space. The fourth dimension
represents evolution in imaginary time and its extent is $\hbar/k_BT$ where $\hbar$ is the the reduced Planck's constant, $k_B$ is the Boltzmann constant and $T$ is temperature.
This system of world lines is simulated efficiently by the quantum Monte Carlo  ``worm algorithm'' \cite{worm}.
Energy gaps for vacancy and interstitial excitations are readily obtained from the
exponential decay in imaginary time of the one-body Green's function of the system~\cite{Boninsegni06}.

\begin{figure}
\centerline{\includegraphics[angle = 0, width=0.9\columnwidth]{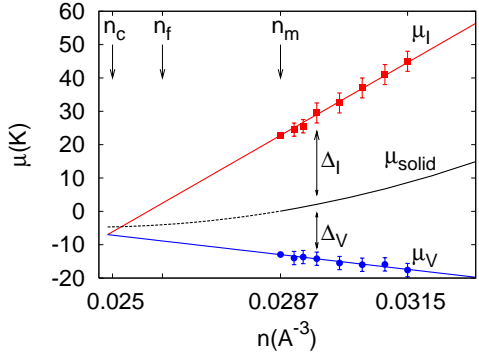}}
\caption{(Color online)  Extrapolation of vacancy ($\Delta_V$) and interstitial ($\Delta_I$) gaps shows that the density corresponding to closing the gap lies in the liquid phase.
The data points as a function of density for the ideal hcp solid are taken from Ref.~\onlinecite{Boninsegni06}.  By $\mu_V$($\mu_I$) we denote
the threshold chemical potentials for doping the system with vacancies(interstitials).
}
\label{fig:hydrostatic}
\end{figure}

\paragraph{\bf Hydrostatic decompression}
After initial attempts to detect the supersolid state in \he4 failed, it was suggested that 
a metastable supersolid can form in crystals decompressed below their
melting density of $n_m = 0.0287$ \Am3 (the freezing density is $n_f = 0.0261$ \Am3 ). 
The idea turned out to be impossible to realize
experimentally, and now we understand why it was implausible in the first place. In
Fig.~\ref{fig:hydrostatic} we show the density dependence of the vacancy and interstitial
gaps. Though the simulation data are available for densities $n> n_m$ they can be
readily and reliably extrapolated to lower densities $n<n_m$ using the near perfect
linear density dependence. One can easily see that metastable hcp crystals remain
insulating all the way to liquid densities and even beyond. It it
unlikely that the solid structure will survive for long at liquid densities, and neither will the possible supersolid phase at density $n_c\approx 0.025$ \Am3. The hydrostatic strain
required to reach this density is about $(n_m/n_c-1)^{1/3} \approx 13.5~\%$.
Assuming solid compressibility at melting \cite{Jarvis} the required underpressure 
is close to -25 bar.


The stability of the metastable supersolid phase depends on the sign of the
effective interactions between vacancies since an attractive Bose gas is unstable:
for the vacancy gas this means a collapse to the lower density liquid.
To determine the sign and strength of the vacancy-vacancy interaction one has
to know their mass $m^{*}$ and their pair correlation function $\nu(r)$.
In the ideal gas $\nu(r)$ is enhanced at short distances by a factor of two relative
to the large distance limit. Correspondingly, for repulsive/attractive interactions
$\nu(r)$ is suppressed/enhanced relative to the ideal gas behavior.
If attractive interactions are so strong that vacancies actually form a bound state, then
$\nu(r)$ is enhanced exponentially, and the decay of the correlation function with distance
can be used to determine the binding energy from $\nu(r) \propto \exp (-2\sqrt{m^{*} E_b} r)$.

\begin{figure}
\includegraphics[angle = 0, width = 0.9\columnwidth]{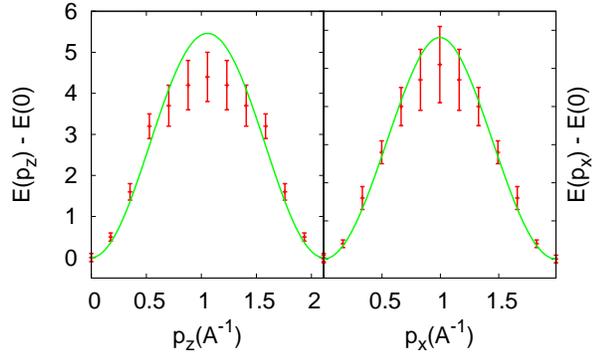}
\caption{(Color online) Vacancy dispersion relation of the lowest band along the reciprocal lattice directions $p_z$ and $p_x$ for a system of $12 \times 12 \times 12$ particles calculated at a density $n = 0.0292$\Am3 and temperature $T = 0.2$K.
For the hcp lattice with two atoms in the unit cell there are two hopping amplitudes, one in the basal plane  ($t_{\perp}$),  and one ($t_z$),
along the $\Gamma A$ direction.   
At low temperatures the Monte Carlo method will project out the lowest branch of the dispersion relation.
}
\label{fig:disp}
\end{figure}

In Fig.~\ref{fig:disp} we present the vacancy dispersion relation which is
analyzed within the tight-binding approximation of Ref.~\cite{Galli03}, obtaining  tunneling amplitudes $t_z = 0.45(5)~K$ and $t_{\perp} = 0.50(5)~K$ and
effective masses $m_{z}^* = 0.45(5)$ and $m_{\perp}^* = 0.42(5)$ in units of the bare \he4 mass.
We find that the effective mass is about two times
lighter than that of \he4 atoms, and nearly isotropic.
There is a small difference with the variational calculation of Ref.~\cite{Galli03} which reported
$m_{\perp}^* = 0.31$ in the basal plane and $m_z^* = 0.38$ along the $\Gamma A$ direction.

\begin{figure}
\includegraphics[angle = 0, width=0.9\columnwidth]{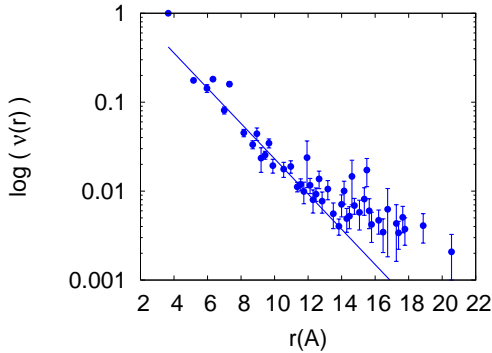}
\caption{(Color online) The probability of finding two vacancies at a distance $r$.
The initial decay is fit with the exponential function $\nu(r) \propto \exp (-2\sqrt{m^{*} E_b} r)$ with $E_b=1.4(5)$K.
The simulation was performed for the hcp solid with $6 \times 6 \times 6$ lattice points and $214$ atoms
at a density $n = 0.0292$\Am3~ and temperature $T = 0.5$K. At large distances the probability reaches a plateau due to finite temperature effects.
}
\label{fig:correlations}
\end{figure}

In Fig.~\ref{fig:correlations} we show the vacancy-vacancy correlation function, decreasing exponentially with distance. This is concrete evidence showing that interactions between vacancies are
attractive and strong enough to form a bound state. From the exponential fit we estimate the binding
energy to be $E_b=1.4(5)$~K. We conclude that the hcp structure at $n_c$ is kinetically unstable
against collapse to the liquid state, and a supersolid in a decompressed crystals
is not possible in the homegeneous setup. As a side remark we note that the equilibrium density
of vacancies at finite temperature $n_V = (m^*T/2 \pi)^{3/2} e^{- \Delta_V/T} \approx 0.015\:n\:e^{- \Delta_{\rm V}/T}$ involves a very small pre-exponential factor 
in addition to the exponential suppression, which rules out their possible role
in thermodynamics at temperatures below $1K$.

\begin{figure}
\centerline{\includegraphics[width=0.9\columnwidth]{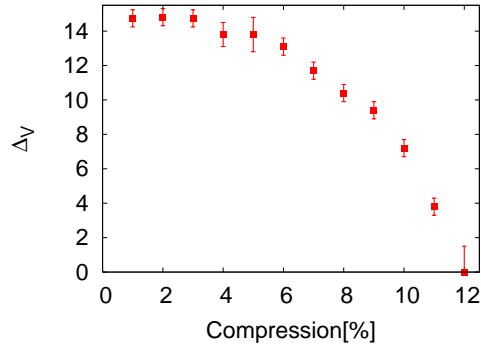}}
\caption{(Color online) Gaps for vacancy formation under anisotropic compression. An ideal hcp solid of size $6 \times 8 \times 8$ at density $n = 0.0292$ \Am3 is subjected to anisotropic compression along the $z$ direction.
The volume is kept constant by dilatation in the basal plane. The temperature is $T = 0.25 {\rm K} $.
}
\label{fig:anisotropic}
\end{figure}

\paragraph{\bf Anisotropic stress}  We now turn to the study of crystals with non-zero
diagonal components of the strain tensor $u_{xx},\, u_{yy}$ and $u_{zz}=-u_{xx}-u_{yy}$.
The phenomenology of the strain induced supersolidity in the hcp structure \cite{Dorsey06}
is based on the minimal energy density allowed by symmetry:
$\epsilon = (a_z-a_\perp)u_{zz}|\psi|^2$ for the anisotropic stress and $\epsilon = (a_zu_{zz}+a_\perp (u_{xx}+u_{yy}))|\psi|^2$ for hydrostatic stress, where $\psi$ is the superfluid order parameter.
In the simulations we find ( see Fig.~\ref{fig:anisotropic})
that a strain $u_{zz}=2u_{xx}=2u_{yy}$ of about 10-12\% is necessary to close the gap for vacancy
formation, and at higher strain the hcp structure collapses. Such a strain corresponds
to a stress  $\sigma_{zz}=C_{33}u_{zz} + C_{13}(u_{xx}+u_{yy})=(C_{33}-C_{13}) u_{zz}$ of  approximately 50 bar,
hardly achievable under realistic experimental conditions. Our data shows that within error bars there is initially no dependence of the gap on anisotropic compression. Thus
the linear coupling to the anisotropic strain is close to zero, $a_z \approx a_\perp$, 
and one has to go beyond linear theory to account for the observed effects, and the closing of the gap.

\begin{figure}
\centerline{\includegraphics[width=0.9\columnwidth]{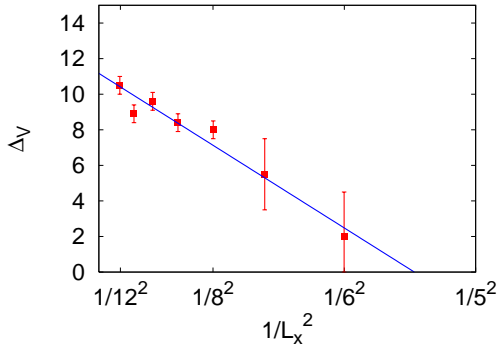}}
\caption{(Color online) Gaps for vacancy formation under shear stress.
An ideal hcp solid of size $L_x \times 12 \times 12$ at density $n = 0.0292 $\Am3 is subjected to shear stress along the $z$ direction by uniformly deforming the sample so that the atoms at boundary
$x=a L_x$ (with $a=3.645$\AA) are shifted exactly by one lattice period $c=\sqrt{8/3}a$ along the c-axis in order to match the periodic boundary conditions. The strain introduced is defined by $u_{zx}=\sqrt{2/3}/L_x$ and, for the chosen sizes $L_x=12-5$, the strain  ranges from $0.07$ down to $0.15$.
The temperature is $T = 0.2 {\rm K}$.
}
\label{fig:shearstress}
\end{figure}

\paragraph{\bf Shear stress} By symmetry there is no linear coupling
between  $| \psi |^2$ and shear strain characterized by the off-diagonal components
$u_{zx}, \, u_{yx},\, u_{zy}$. The anticipated strain dependence of the gap is quadratic,
$\Delta_V = \Delta_V^{(0)} [1- (u_{zx}^2 + u^2_{zy}/u_c^2]$. Our results in Fig.~\ref{fig:shearstress}
allow us to estimate the critical value of shear strain as $u_c\approx 0.15$. 
Using measured values of the elastic modulus $C_{44}\approx$ 120-130 bar \cite{Greywall77b}
the corresponding critical shear stress, $2C_{44}u_{c}$, for closing an insulating gap in 
the hcp \he4 is about 35 bar.

\paragraph{\bf Superfluidity along crystalline defects}
Even though the homogeneous strain-induced supersolid phase is unstable, it can form
locally if non-uniform strain accumulating close to structural defects exceeds the critical value
and destabilizes parts of the crystal. The most probable candidates for such highly strained
superfluid defects are dislocations~\cite{Schevchenko, Dorsey06,Toner08}, with
edge dislocations attracting most attention in the past because they produce strain coupling linearly
to the superfluid order parameter. Contrary to expectations, the first evidence for superfluidity in the
dislocation core was reported for screw dislocations oriented along the $\hat{z}$-direction \cite{Boninsegni07}.
These defects are characterized by the non-zero values of $u_{zx}$ which can be estimated
by dividing the modulus of the Burgers vector $b_z=\sqrt{8/3}a$ by twice the circumference of the
circle going through the atoms closest to the core, $4 \pi a/\sqrt{3}$. The estimated
strain $\sqrt{u^2_{xz} + u^2_{yz}}=1/(\sqrt{2}\pi) \approx 0.22$ exceeds the threshold value
of $u_c=0.15$ found above. This explains why the superfluid density of the screw dislocation
involves nearly all atoms closest to the nucleus.

An alternative mechanism to release large strain at the core is to split the core into two parts, each characterized by half the Burgers vector. Whether this actually happens depends on the dislocation
type and the energy of the domain wall stretching between the cores (the energy of stacking faults
in hcp lattices is relatively low). For screw dislocations the core splitting mechanism is
not effective.

Similar considerations apply to edge dislocations and grain boundaries, where simulations have detected superfluidity in some randomly oriented grain boundaries \cite{Pollet07}, but the majority of
small angle boundaries and stacking faults were found to be insulating. Simple estimates of strain in the vicinity of the defect cores range from 0.05 to 0.2, depending on the geometry and an effective Burgers vector, which may be reduced if the split-core mechanism is effective.

\paragraph{\bf Conclusions}
There is growing experimental evidence for
unexpected properties of solid \he4 at low temperature, including superfluid mass flow through
the solid. The observed dependence on sample history, growth conditions, annealing and cooling procedures, indicate that crystalline defects are important for our understanding of the most quantum solid in Nature. The actual structure of defects is essentially unexplored territory, not less interesting than the solid matrix they reside in. Similar exciting phenomena occur at defects and interfaces in electronic systems~\cite{Nabutovski, Dagotto}. In this work we determined
the critical values of the strain which are required to destabilize the hcp structure of \he4 by closing
its insulating gap, and find that these thresholds are small enough to be exceeded at the dislocation cores and grain boundaries. Local stresses thus hold the key to supersolidity of \he4.


This work was supported by the National Science Foundation
under Grants Nos. PHY-0653183 and PHY-065135, CDRF grant No. 2853,
the Natural Science and Engineering Research Council of Canada 
under  research grant G121210893, and the Swiss National Science Foundation. 
Simulations were performed on Hreidar (ETH Zurich), Typhon and Athena (CSI), and Masha (UMass) Beowulf clusters.

\end{document}